# A Simple View on Large-Signal Resonant-Tunneling-Diode Dynamics

Petr Ourednik[1,*], Dinh Tuan Nguyen[1], and Michael Feiginov[1]
[1]Department of Electrical Engineering and Information Technology, TU Wien, Vienna 1040, Austria
*)petr.ourednik@tuwien.ac.at

*Abstract*— We present a model for an accurate description of the large-signal resonant-tunneling-diode (RTD) dynamics, which allows for a simple and intuitive analysis in terms of dynamical trajectories in a phase space. We show that the RTD admittance can be accurately described by a simple RLRC equivalent circuit, which has a universal configuration, but with different circuit parameters in the large- and small-signal cases.

## I. INTRODUCTION

RESONANT-TUNNELING diodes (RTDs) are the fastest active semiconductor electronic devices. They exhibit negative differential conductance (NDC) and can be used in sub-THz and THz oscillators [1,2]. Recently, a record oscillation frequency of 1.98 THz and an output power of 10 mW at 450 GHz were reported [3,4]. The RTD oscillators' output power depends on RTDs' dynamical large-signal (LS) parameters [5]. However, the LS dynamic modeling of RTDs remains a difficult and cumbersome problem. In the past, extrapolation of the quasi-static RTD characteristics [1,6] or very high-frequency asymptotical models [7] were used for a high-frequency nonlinear analysis. However, most state-of-the-art oscillators operate at intermediate frequencies where the above models are questionable, and an accurate account of charge-relaxation processes in RTDs is required. In this paper, we present a general model capable of describing the full-frequency span of LS RTD operation. The calculation details are presented in [8]; here, we show that our model provides a simple and intuitive insight into the LS RTD dynamics.

## II. LS RTD DYNAMICS

We start the analysis of LS RTD dynamics with a self-consistent solution of the Schrödinger and Poisson equations. We note that the RTD band profile, tunnel probabilities, and internal current densities can be uniquely defined by just two parameters: RTD bias ($V_\text{RTD}$) and the charge density ($N_\text{2D}$) in the RTD quantum well (QW). Those parameters define in a natural way a phase space (as the RTD-state variables) for the analysis of LS RTD dynamics. Further, we calculate a map of $\Delta j \stackrel{\text{def}}{=} j_\text{e} - j_\text{c}$ in this phase space, where $j_\text{e}$ and $j_\text{c}$ are the RTD emitter and collector current densities, respectively, see a map of $\Delta j$ in Fig. 1. The RTD dynamics can be then described by a single charge-continuity equation:

$$\partial_t e N_\text{2D} = \Delta j(V_\text{RTD}, N_\text{2D}),  \qquad (1)$$

where $e$ is the elementary electron charge. For a given externally-applied arbitrary-time-dependent $V_\text{RTD}$, Eq. (1) determines $N_\text{2D}$ trajectories in the phase space, which fully define the LS RTD dynamics. The condition $\Delta j = 0$ determines the steady states of the system, corresponding to $N_\text{2D}^\text{DC}(V_\text{RTD})$ on the DC RTD I-V curve, see Fig. 1.

Further, one can show (Poisson and charge-continuity equations) that knowing the dynamical $N_\text{2D}$ trajectories, the external RTD current ($j_\text{RTD}$) can be calculated as:

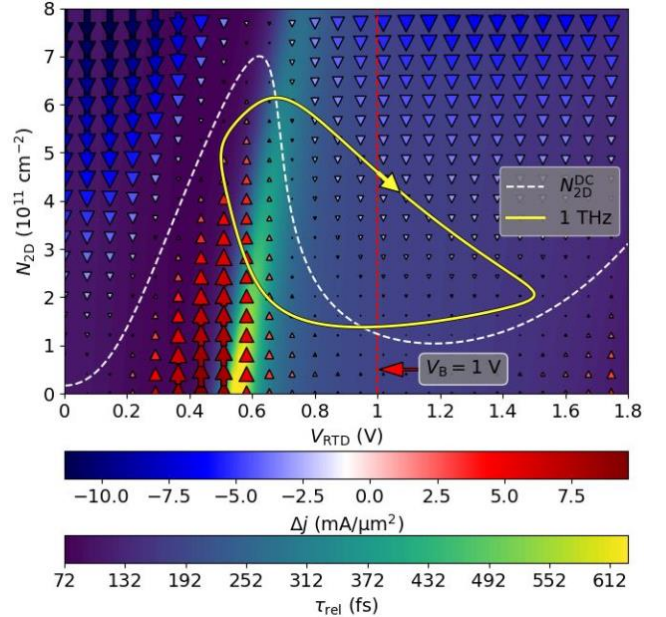

**Fig. 1.** A color map of differential relaxation time constant ($\tau_\text{rel}$) and an arrow/color map of $\Delta j$ in the phase space $V_\text{RTD}$ and $N_\text{2D}$ for an example RTD, studied in [6,8]. The white dashed line shows stationary $N_\text{2D}^\text{DC}(V_\text{RTD})$, corresponding to the DC I-V curve. The corresponding I-V is shown in Fig. 2 a). The yellow line shows an example dynamical trajectory at 1 THz with an AC-voltage amplitude of 0.5 V.

$$j_\text{RTD} = e N_\text{2D} \nu_\text{c} + \frac{C_\text{ec}}{C_\text{ew}} \partial_t e N_\text{2D} + C_\text{ec} \partial_t V_\text{RTD},  \qquad (2)$$

where $\nu_\text{c}$ is the electron tunneling rate through the RTD collector barrier, $C_\text{ec}$ is the RTD geometrical capacitance, and $C_\text{ew}$ is the capacitance between the emitter and the QW.

As a next step, we linearize Eq. (1) as:

$$\partial_t \delta e N_\text{2D} = \frac{\partial \Delta j}{\partial e N_\text{2D}} \delta e N_\text{2D} = -\frac{1}{\tau_\text{rel}} \delta e N_\text{2D},  \qquad (3)$$

where $\delta$ denotes small deviations and $\tau_\text{rel}$ is a differential relaxation time [9] at every point in the phase space, a map of $\tau_\text{rel}$ is shown in Fig. 1. Around $N_\text{2D}^\text{DC}(V_\text{RTD})$, $\tau_\text{rel}$ corresponds to a small-signal (SS) RTD relaxation time [9].

This quite intuitive picture provides a qualitative understanding of the LS RTD dynamics, e.g., the map of $\tau_\text{rel}$ gives an estimate for the time scales occurring in the dynamical problem. Depending on the operating range of the driving voltage $V_\text{RTD}$, one can see the relevant values of $\tau_\text{rel}$ on the map.

Even more, integrating $1/\tau_\text{rel}$ along a relaxation trajectory for a fixed $V_\text{RTD}$, we can define a LS relaxation time for arbitrary perturbations of $N_\text{2D}$, corresponding to a relaxation process bringing this perturbed $N_\text{2D}$ to the stationary (DC) $N_\text{2D}^\text{DC}(V_\text{RTD})$.

Further, we apply to the RTD a sinusoidal driving voltage: $V_\text{RTD} = V_\text{B} + V_\text{AC} \cos(\omega t)$, where $V_\text{AC}$ is the AC amplitude, $V_\text{B}$ is the DC bias, and $\omega$ is the angular frequency. That is a typical operation regime of an RTD in an oscillator. Qualitatively, when the operating frequency is small compared to the rate of the relaxation processes, $N_{2D}$ will have enough time to relax to $N_{2D}^{DC}(V_{RTD})$ at every bias point, and the trajectory closely follows the $N_{2D}^{DC}(V_{RTD})$ curve (quasi-static regime). In the opposite case of very high frequencies, $N_{2D}$ will remain nearly constant since the relaxation cannot occur, and the trajectory will be nearly horizontal in this case, see [7]. In an intermediate case (around 1 THz in our example), the dynamical trajectory will look like that shown in Fig. 1; the relaxation processes will play an essential role in the RTD dynamics in this case.

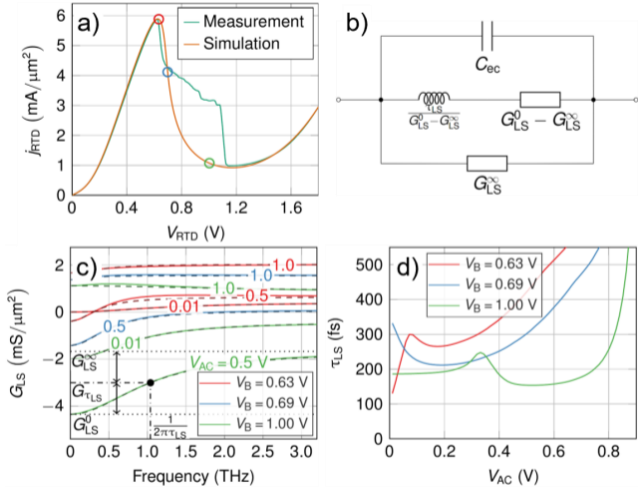

**Fig. 2.** a) I-V curve of the RTD studied in [6,8]. b) LS equivalent RLRC circuit of the RTD. c) frequency dependence of the LS conductance of the studied RTD for bias points marked in a) and for several values of $V_{AC}$. The dashed lines correspond to the fitting curves with equivalent circuit in b). d) LS time constant dependence on the AC voltage amplitude.

As a mechanical analogy for our RTD system, we can consider a record player with a vinyl plate. When the plate is rotating slowly, the needle of the record player will follow the grooves of the vinyl plate and stays in the valley (quasi-static trajectory). However, playing the record faster, the needle is unable to stay in a minimum of a groove, and it will follow a dynamic trajectory around the minimum. At very high speed, the needle will be jumping over the hills. This analogy holds well as long as the dynamics of the needle are dominated by friction and the inertia of the needle holder is negligible.

### III. LS RTD Admittance

Apart from an intuitive illustration of the RTD dynamics, the maps of $\Delta j$ and $\tau_{rel}$ in the phase space provide a means of accurate quantitative analysis of the dynamic trajectories and LS RTD response.

We apply a sinusoidal driving voltage to RTD, as above, calculate the dynamic trajectories, and then use Eq. (2) to compute the external RTD current. Evaluating the current's first harmonic, we get the LS admittance, whereas the real part is shown in Fig. 2 c) for a pick of bias points and amplitudes for an example RTD measured and analyzed in [6,8]. The roll-off profile of the RTD conductance can be accurately approximated by a simple model derived in the past for the SS analysis [9]. The model is shown in Fig. 2 b) and reads for the LS RTD admittance as:

$$Y_{LS} = G_{LS} + jB_{LS} = j\omega C_{ec}^{(I)} + G_{LS}^{\infty} + \frac{G_{LS}^0 - G_{LS}^{\infty}}{1+j\omega\tau_{LS}}, \quad (4)$$

where, $G_{LS}$ and $B_{LS}$ are the LS conductance and susceptance, respectively, $G_{LS}^0$ is the DC LS conductance, $G_{LS}^{\infty}$ is the asymptotic high-frequency LS conductance, and $\tau_{LS}$ is a LS time constant. The model's parameters are determined by fitting to the simulated admittance curves, see Fig. 2 c).

Fig. 2 d) depicts the dependency of $\tau_{LS}$ on $V_{AC}$ for the pick of biases. For small amplitudes, $\tau_{LS}$ converges to SS $\tau_{rel}$. For increasing amplitudes $\tau_{LS}$ dramatically changes. The plot shows that for $V_B = 0.69$ V, where NDC is the highest and RTD is the slowest, $\tau_{LS}$ significantly drops with increasing AC amplitude. This is due to the faster relaxation of $N_{2D}$ when we drive the voltage outside of the slowest region.

The output power of RTD oscillators typically increases with the increase of the bias voltage towards to valley region of the I-V curve, making these bias points valuable to analyze. For $V_B = 1$ V, we can observe that $\tau_{LS}$ first increases with $V_{AC}$ and peaks when the voltage sweep culminates in the slow region with the highest NDC. Increasing the AC amplitude further decreases $\tau_{LS}$ even to values below SS $\tau_{rel}$.

### IV. Summary

We show that the LS RTD dynamics can be intuitively described with the $\Delta j$ and $\tau_{rel}$ maps in the $V_{RTD}$ and $N_{2D}$ phase space. This approach also provides an accurate quantitative description of the dynamical LS RTD response. Further, the LS RTD admittance can be accurately represented by a simple equivalent circuit, which has a universal configuration, but with LS circuit parameters deviating from that in the linear SS case. From the LS analysis, we show that the LS time constant of the RTD conductance can be larger or smaller than its SS equivalent. For the RTD oscillators the shorter LS time constant can lead to higher output powers at intermediate frequencies.

### Acknowledgment

This work was supported by FWF project P30892-N30.

### References

[1]. M. Asada and S. Suzuki, "Terahertz emitter using resonant-tunneling diode and applications," *Sensors*, vol. 21, no. 4, p. 1384, 2021.
[2]. M. Feiginov, "Frequency limitations of resonant-tunnelling diodes in sub-THz and THz oscillators and detectors," *Journal of Infrared, Millimeter, and Terahertz Waves*, vol. 40, no. 4, pp. 365–394, 2019.
[3]. R. Izumi, S. Suzuki and M. Asada, "1.98 THz resonant-tunneling-diode oscillator with reduced conduction loss by thick antenna electrode," *2017 42nd International Conference on Infrared, Millimeter, and Terahertz Waves (IRMMW-THz)*, pp. 1-2, Cancun, Mexico, 2017.
[4]. Y. Koyama et al., "A High-Power Terahertz Source Over 10 mW at 0.45 THz Using an Active Antenna Array With Integrated Patch Antennas and Resonant-Tunneling Diodes," *IEEE Transactions on Terahertz Science and Technology*, vol. 12, no. 5, pp. 510-519, Sept. 2022.
[5]. C. Spudat, P. Ourednik, G. Picco, D. T. Nguyen and M. Feiginov, "Limitations of Output Power and Efficiency of Simple Resonant-Tunneling-Diode Oscillators," *IEEE Transactions on Terahertz Science and Technology*, vol. 13, no. 1, pp. 82-92, Jan. 2023.
[6]. P. Ourednik, T. Hackl, C. Spudat, D. Tuan Nguyen, and M. Feiginov, "Double-resonant-tunneling-diode patch-antenna oscillators," *Applied Physics Letters*, vol. 119, no. 26, p. 263509, 2021.
[7]. M. Feiginov, C. Sydlo, O. Cojocari, and P. Meissner, "High-frequency nonlinear characteristics of resonant-tunnelling diodes," *Applied Physics Letters*, vol. 99, no. 13, p. 133501, 2011.
[8]. P. Ourednik, G. Picco, D. Tuan Nguyen, C. Spudat, and M. Feiginov, "Large-signal dynamics of resonant-tunneling diodes," *Journal of Applied Physics*, vol. 133, no. 1, p. 014501, 2023.
[9]. M. N. Feiginov, "Does the quasibound-state lifetime restrict the high-frequency operation of resonant-tunnelling diodes?," *Nanotechnology*, vol. 11, no. 4, pp. 359–364, 2000.